\documentclass[aps,prl,twocolumn,showpacs,superscriptaddress,groupedaddress]{revtex4-1}

\usepackage{hyperref}
\usepackage{multirow}
\usepackage{mathrsfs,amsmath}
\usepackage{verbatim}
\usepackage{amsmath}
\usepackage{amsfonts}
\usepackage{amssymb}
\usepackage{amsthm}
\usepackage{bm}
\usepackage{xparse}
\usepackage{graphicx}
\usepackage{xcolor}
\usepackage{textcase}
\usepackage{url}
\usepackage{lipsum}
\usepackage{appendix}
\usepackage{dsfont}
\usepackage{epstopdf}
\usepackage{footnote}

\newcommand{\ket}[1]{| {#1} \rangle} 
\newcommand{\bra}[1]{\langle {#1} |} 
\DeclareDocumentCommand{\Tr}{m m O{\big}}{{\rm Tr}_{\:\!{#1}}#3({#2}#3)}

\begin{document}
\title{Coherence Equality and Communication in a Quantum Superposition}
\author{Flavio Del Santo}
\affiliation{Vienna Center for Quantum Science and Technology (VCQ), Faculty of Physics,
Boltzmanngasse 5, University of Vienna, Vienna A-1090, Austria
}
\affiliation{
Institute for Quantum Optics and Quantum Information (IQOQI),
Austrian Academy of Sciences, Boltzmanngasse 3,
A-1090 Vienna, Austria}
\author{Borivoje Daki\'c}
\affiliation{Vienna Center for Quantum Science and Technology (VCQ), Faculty of Physics,
Boltzmanngasse 5, University of Vienna, Vienna A-1090, Austria
}
\affiliation{
Institute for Quantum Optics and Quantum Information (IQOQI),
Austrian Academy of Sciences, Boltzmanngasse 3,
A-1090 Vienna, Austria}

\date{\today}

\begin{abstract}
In this paper, we introduce a ``coherence equality'' that is satisfied by any classical communication --i.e. conveyed by a localized carrier traveling along well defined directions. In contrast, this equality is violated when the carrier is prepared in coherent quantum superposition of communication directions. This is phrased in terms of the success probability of a certain communication task, which results to be always constant and equal to $1/2$ in the classical case. On the other hand, we develop two simple quantum schemes that deviate systematically from the classical value, thus violating the coherence equality. Such a violation can also be exploited as an operational way to witness spatial quantum superpositions without requiring to recombine the modes in a standard interferometer, but only by means of spatially separated local measurements.
\end{abstract}

\maketitle
\section{Introduction}
Quantum superposition principle states that an arbitrary linear combination of two physical states is still a valid quantum state. Such a principle lies at the core of genuine quantum behaviors. In fact, it even supervenes quantum entanglement, which can be regarded as a particular state of superposition that combines two or more joint degrees of freedom.
Since the early days of quantum theory, physicists have been using the effect of quantum superposition (or coherence) as the foremost observable evidence to discriminate between the classical and the quantum domains. It has been also a prime conceptual tool for testing the limits of quantum mechanics, like in the notorious Schr{\"o}dinger's cat gedankenexperiment \cite{schrodinger1935gegenwartige}.

In more recent years, effects based on coherence played a central role in the revolutions of quantum information and quantum technologies, allowing a plethora of novel achievements that are fundamentally unattainable in classical scenarios, such as secure communication \cite{bennett2014quantum}, and algorithms with an exponential advantage over their classical counterpart \cite{deutsch1992rapid, shor1994algorithms}. Moreover, it was shown that quantum superposition can be used as a resource for quantum communication \cite{feix2015quantum,jia2019causal,guerin2016exponential,chiribella2013quantum, chiribella2018indefinite,Massa18,del2018two} and can lead to novel effects such as the enhancement of a classical channel capacity \cite{chiribella2018indefinite}, secure anonymous communication \cite{Massa18} or the doubling of the bandwidth of a classical channel \cite{del2018two}.

However, it is well known that the direct observation of a quantum superposition of distinct states is not possible. For example, in the celebrated double-slit experiment, if one detects the presence of the particle at either of the two slits, no quantum effects are manifested, and even microscopic particles (e.g., electrons) resemble classical bullets. 
Accordingly, witnessing a quantum superposition requires a particular type of indirect observation, usually achieved by spatially separating the wave function before recombining them in an interference experiment.
Consider, for instance, the aforementioned double-slit experiment, where each slit has the option to be open or closed, labeled  by ``0'' and ``1'', respectively.  This leads to four possible configurations: ``00'', ``01'', ``10'', or ``11'',  where the position of the digit indicates the first or the second slit, respectively. The ``non-classicality'', i.e. the presence of a quantum superposition, can then be measured by the interference term \cite{dakic2014density}:
\begin{equation}\label{interf}
I:= \sum_{i,j=0}^1 (-1)^{i\oplus j} p_{ij},
\end{equation}
where $p_{ij}$ is the conditional probability to find the particle on the observation screen, given the configuration $ij$. In a classical scenario $I=0$ always, whereas a non-zero value would represent a witness of a coherent superposition. Exactly the same condition one finds in the case of a Mach-Zehnder interferometer (the simplest instantiation of the double-slit experiment), where a single particle is separated into two paths by a beam-splitter and interference fringes appear at the detector when $I\neq0$. However, as in every other interferometric experiment, it is necessary to recombine the paths at a second beam-splitter (and to introduce a relative phase between the two paths) to observe the interference.

In this paper, we propose an operational way to witness a superposition only using local measurements conducted at separated locations, without the necessity of recombining the paths as in a standard interferometer.
To achieve this, we derive a ``coherence equality'', which is satisfied by any classical resource, but that can be violated by systems that exhibit coherence (i.e., in quantum superposition). We will phrase the problem using the modern language of information and communication tasks that present a systematic difference in the probabilities of success with their classical counterpart, when quantum resources are employed.
\section{Coherence equality and interference}
In this section, we derive a ``coherence equality'', whose violation can be regarded as an operational procedure to witness quantum superposition without the use of a standard interferometer. That is, without the necessity of recombining the two arms to detect an interference pattern (thus effectively using a setup which is only half of a Mach-Zehnder interferometer). 

\begin{figure}[h!]
\centering
\includegraphics[width=8.5cm]{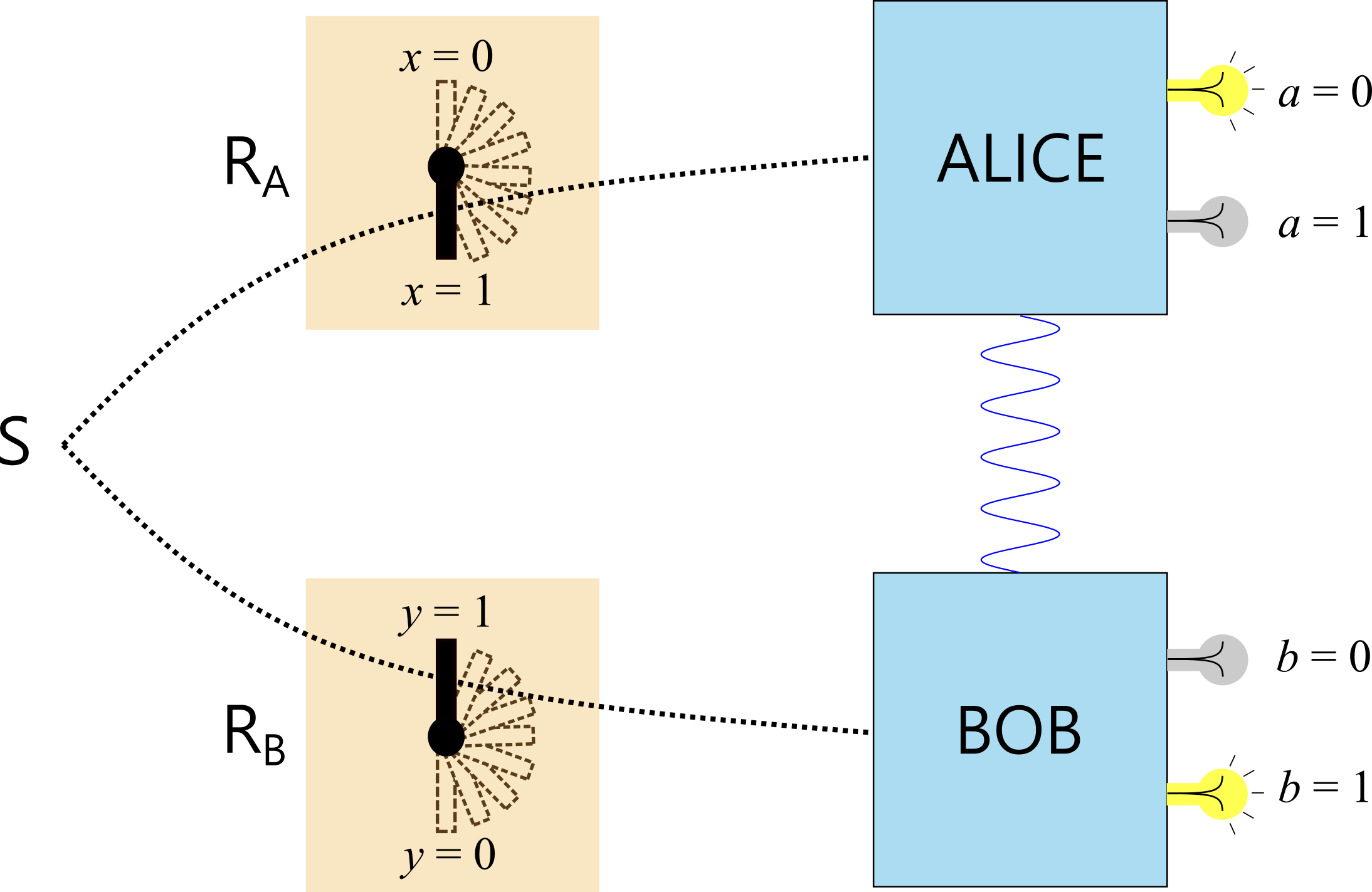}
\caption{\textbf{Scheme of the  ``coherence without re-interference'' communication game}. A source $S$ produces a single ``information carrier'' which can be sent to two parties, Alice and Bob. The latter can share some resources (represented by the blue wavy line). $A$ and $B$ are asked by two referees to output one bit each ($a$ and $b$, respectively) to fulfill a certain task. Movable ``blockers'' are situated along the channels $S - A$ and $S -B$ and their configurations, open or closed, function as encoded inputs, $x$ on Alice's side, and $y$ on Bob's. Quantum mechanics allows a violation of a fundamental classical bound to the probability of success (see main text).}
\label{setup}
\end{figure}

Let us start by considering a scenario like the one depicted in Fig.~\ref{setup}. A source $S$  produces a \emph{single information carrier} (e.g., a single particle) in order to covey a piece of information to Alice ($A$) or Bob ($B$),  who are separated at two different locations. (As we shall see, even if communication is allowed between the two agents --contrarily to non-local scenarios wherein space-like separation is enforced--  the result would still hold unchanged).

The information to be communicated is a bit which is encoded along either of  the paths traveled by the information carrier: $x\in\{0, 1\}$ on the branch leading to Alice, and $y\in\{0, 1\}$ in the branch leading to Bob. The operation of encoding consists in selecting one of the two configurations, open or closed, of a movable ``blocker'' (i.e., an ideally impenetrable barrier) on each of the two channels connecting $S$ to $A$ and $S$ to $B$, respectively. We denote by $x=0$ ($x=1$) the configurations in which the blocker is open (closed) along the channel $S-A$; in the same fashion, the input $y=0$  ($y=1$) means that the blocker in channel $S-B$ is open (closed). No information can reach the receiver if a blocker is placed in the respective path (i.e., if $x=1$ and/or $y=1$).

The parties $A$ and $B$ then (potentially) receive the information carrier and they perform local measurements whose binary outputs are labeled by $a\in\{0, 1\}$ and $b\in\{0, 1\}$, respectively. Moreover, $A$ and $B$ are allowed to share some resources such as classical shared randomness, entanglement or coherence. {Since in the classical case the information carrier is a well-localized object at any instant in time (which means that the particle will take only one definite path among $S-A$ and $S-B$) the inputs $x$ and $y$ can causally influence the outputs $a$ and $b$ in two mutually exclusive, possible ways (see Fig.~\ref{causal}).
\begin{figure}[h!]
\centering
\includegraphics[width=5.5cm]{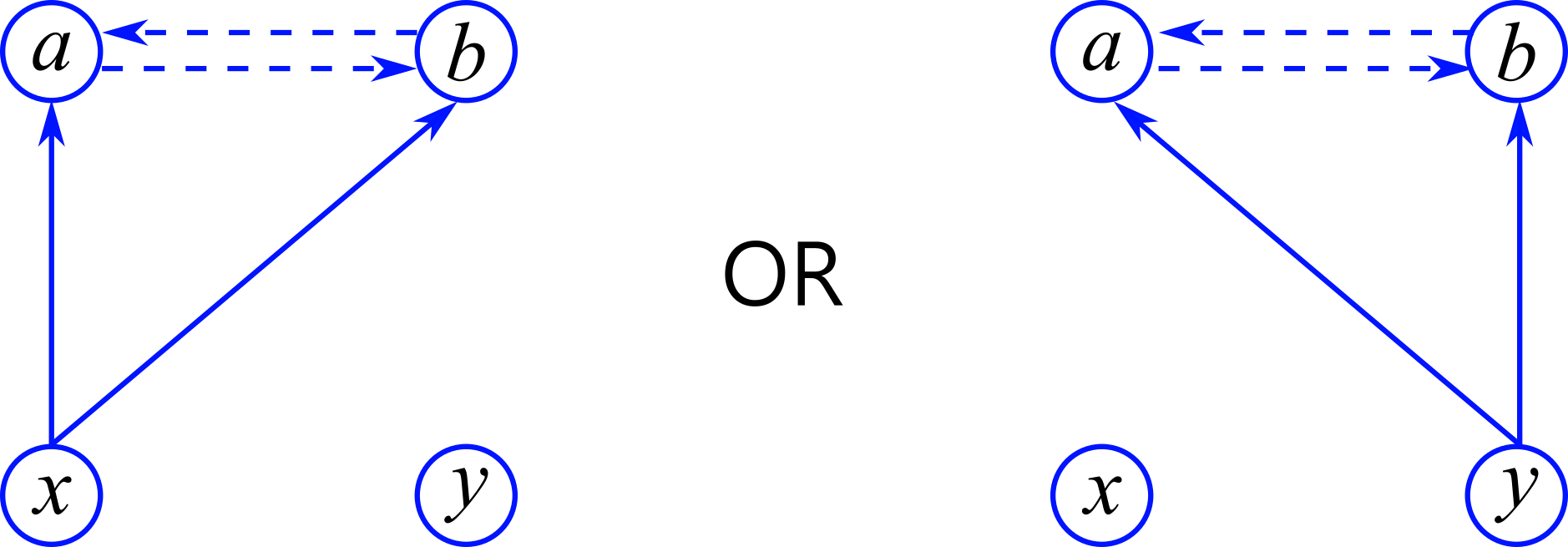}
\caption{\textbf{Causal diagram}. Classically, a single information carrier can convey information only in one of two depicted scenarios. Either $x$  or $y$ influences $a$ and $b$. {The dashed arrows indicate that, in principle, there could be communication between $a$ and $b$}.
}
\label{causal}
\end{figure}
Therefore, the joint probability distribution of all the outputs, given the inputs, is a classical mixture of the distributions corresponding to one-way signaling distributions (either $S$ communicates to $A$, or to $B$):
\begin{equation}
p(ab|xy)=\lambda_{S-A} p_{S-A}(ab|x)+\lambda_{S-B} p_{S-B}(ab|y)
,
\label{class}
\end{equation}
where $\lambda_{S-A}$ and $\lambda_{S-B}$ are non-negative constants that add up to the unity.}
To measure interference effects, in analogy with Eq.~\eqref{interf}, we define:
\begin{equation}\label{Iquant}
I_{ab}= \sum_{x,y=0}^1 (-1)^{x\oplus y} p(ab|xy).
\end{equation}
It follows immediately  from \eqref{class} that for classical systems
\begin{equation}\label{Belleq}
I_{ab}^{Class}=0.
\end{equation}
We call the above expression ``coherence equality'', because any deviation from the value 0 would imply that the information carrier is a non-classical object that exhibit coherence.
{
Note that communication between the parties is in principle allowed, since the derivation of Eq.~\eqref{Belleq} is independent of the separation between $A$ and $B$ (which could indeed be replaced by a single agent). However, for the argument here discussed, we maintain the parties separated at two different locations, because it is our main aim to show that spatially separated agents can certify the presence of quantum coherence only by means of local operations and classical communication, i.e. without closing a standard interferometer. Moreover, such a two-party scenario finds applications in cryptographic protocols \cite{Massa18, massa2019experimental}.
}
\subsection{``Coherence without re-interference'' communication game}
Following a recent trend that aims at quantifying the discrepancy between classical and  quantum scenarios by computing the success probability of quantum XOR non-local games \cite{ambainis2010nonlocal, regev2015quantum,Brunner14}, we define the ``coherence without re-interference'' game, that provides an operational procedure to demarcate the classical resources from the quantum ones. 

Consider again the setup of Fig.~\ref{setup}, but where in this case two referees, $R_A$ and $R_B$, encode inputs $x$ and $y$ (randomly assigned with uniform distribution) by opening or closing their blocker, as described above. The referees challenge the parties $A$ and $B$ to return outputs $a$ and $b$, respectively, which ought to fulfill the following relation:
\begin{equation}
a\oplus b = x\oplus y.
\label{game}
\end{equation}

{
The easiest classical strategy to play the game, would be for Alice and Bob to detect the presence of the particle at their respective positions, each outputting 1 if a particle is detected or 0 otherwise. From the distribution in~\eqref{class}, it is trivial to see that the (non-vanishing) probabilities read
\begin{align}
\nonumber
&p(10|00)= \lambda_{S-A}, ~~ p(01|00)= \lambda_{S-B} \\\nonumber
&p(10|01)= \lambda_{S-A}, ~~ p(00|01)= \lambda_{S-B} \\\nonumber
&p(00|10)= \lambda_{S-A}, ~~ p(01|10)= \lambda_{S-B} \\\nonumber
&p(00|11)= 1.
\end{align}
It follows that the probability of fulfilling relation \eqref{game} is given by $P_{win}=1/4(\lambda_{S-A}+\lambda_{S-B}+1)=1/2$, which satisfies the coherence equality $I_{ab}=0$ in Eq.~\eqref{Iquant}, is satisfied.}

In general, the probability of success, $P_{win}$, is given by the following expression:
 \begin{eqnarray}
P_{win}&=&\frac{1}{4}[p(00|00)+p(00|11)+ p(01|01)+p(01|10) + \nonumber\\ 
&&p(10|01)+p(10|10)+p(11|00)+ p(11|11)]\nonumber\\
&=&\frac{1}{2}+\frac{1}{4}(I_{00}+I_{11}).\nonumber\\
\label{classicalwinprob}
\end{eqnarray}
Since $I_{ab}^{Class}=0$ for all $a$ and $b$, we get $P_{win}^{Class}=\frac{1}{2}$. This means that, using only classical resources, $A$ and $B$ will always achieve the same probability of success of $1/2$, regardless of the strategy they choose. 

{In practice, the probability of deviation of the relative frequency from $1/2$ is exponentially suppressed with the number of experimental trials for any possible classical scenario (see Appendix A).}

\subsection{Quantum task: Communication in quantum superposition}

\textbf{\textit{First example.}}
{Let us consider now the analogous quantum scenario. Let the state of the particle be an equal-weighted superposition of directions of communication, i.e. $\ket{\psi}_S=\frac{1}{\sqrt 2}\left(\ket{A}_S+\ket{B}_S\right)$. The states $\ket A$ and $\ket B$ are taken to mean that the particle is in the path leading towards Alice or Bob, respectively, whereas the subscript $S$ indicates that this particle was created at the source.
Consider now that the internal mechanism of the measuring devices located at the two separate positions, $A$ and $B$, have some pre-shared coherence. Namely, an ancillary identical (i.e. indistinguishable) particle, previously prepared in the superposition state $\ket{\psi}_M=\frac{1}{\sqrt 2}\left(\ket{A}_M+\ket{B}_M\right)$, where the subscript $M$ refers to the measurement device.} The quantum particle produced in $S$ travels from the source to $A$ and $B$ and, after passing through the encoding ports, arrives to the measurement devices where it gets measured together with the particle $M$.

Consider a strategy where the players agree to output a random bit each time they detect two or no particles at their respective locations. This event occurs if both slits are open ($x=y=0$) with probability $1/2$; if one slit is open and the other is closed ($x\oplus y=1$) it occurs with probability $3/4$;  finally, for both slits closed ($x=y=1$) it occurs always. Overall, the player will output random bits in three quarterrs of the runs and corresponds to the classical probability of success of $1/2$.

Let us now consider the cases when one particle is detected by Alice and the other by Bob, which would occur in the other quarter of the runs. Using the formalism of Fock space , we introduce four ladder operators $a^{\dagger}_{S/M}$ and $b^{\dagger}_{S/M}$, which can be either fermionic or bosonic, to designate the two different locations for each of the particles.  For instance, the joint state of the particles (before measurement), for the case of both slits open ($x=y=0$) is given by 
\begin{equation}\label{00state}
\frac{1}{2}(a^{\dagger}_S+b^{\dagger}_S)(a^{\dagger}_M+b^{\dagger}_M)
\ket{0}_{AS}\ket{0}_{AM}\ket{0}_{BS}\ket{0}_{BM}.
\end{equation}
Here, for example, $\ket{0}_{AS}$ labels the vacuum state for the source particle located at  Alice's side. 
{Making explicit the action of the ladder operators on the vacuum states, allows to show the difference between fermionic and bosonic statistics:
\begin{align}
\nonumber
&\ket{1}_{AS}\ket{1}_{AM}\ket{0}_{BS}\ket{0}_{BM} + \ket{1}_{AS}\ket{0}_{AM}\ket{0}_{BS}\ket{1}_{BM} \pm \\ \nonumber
&\ket{0}_{AS}\ket{1}_{AM}\ket{1}_{BS}\ket{0}_{BM} + \ket{0}_{AS}\ket{0}_{AM}\ket{1}_{BS}\ket{1}_{BM},
\end{align}
where the $\pm$ holds for bosons/fermions, respectively.}

It is convenient to introduce the following ``qubit'' states for $A$ and $B$,
\begin{align}
   &\ket{0}_{A}=\ket{1}_{AS}\ket{0}_{AM}, &~~& \ket{1}_{A}=\ket{0}_{AS}\ket{1}_{AM}, \\
   &\ket{0}_{B}=\ket{1}_{BS}\ket{0}_{BM}, &~~& \ket{1}_{B}=\ket{0}_{BS}\ket{1}_{BM}.
\end{align}\label{bases}
These states constitute Alice's and Bob's qubits and can be fully manipulated locally, e.g. by means of linear-optical elements, such as beam-splitters and phase-shifters. { Formally, this means to locally apply transformations of the group $SU(2)$ to these qubit states. To translate any matrix $u \in SU(2)$ in second quantization formalism, firstly one needs to identify its generator (Hamiltonian), i.e. $u=e^{i\hat h}$, with $\hat h_{ij}$ being a $2\times2$ Hermitian matrix. The corresponding Hamiltonian, in the Fock space representation, reads $\hat H=\sum_{ij}h_{ij}a_i^\dagger a_j$. This, in turn, defines the generic transformation of $SU(2)$ in Fock space as $U=e^{i\hat H}$. The transformation $U$ preserves the qubit subspace spanned by $\{\ket{0}_A, \ket{1}_A \}$, and its action simply reduces to the action of $u\in SU(2)$ \cite{gerry2005introductory}.
The same holds for the operators on Bob's side. Note also that these operators take the same form for both fermions and bosons.}

The players will then perform the measurements within these qubit-(sub)spaces, i.e. spanned by $\beta_A=\{\ket{0}_A,\ket{1}_A\}$ and $\beta_B=\{\ket{0}_B,\ket{1}_B\}$. Since we are interested only in cases for which the measurement reveals one particle per party, this situation occurs only when $x=y=0$ or $x\oplus y=1$, and we label the three post-selected states by $\ket{\psi_{xy}}$. A simple calculation shows
\begin{eqnarray}
  \ket{\psi_{00}} &=& \frac{1}{\sqrt2}(\ket{0}_A\ket{1}_B \pm \ket{1}_A\ket{0}_B), \\
  \ket{\psi_{01}} &=& \ket{0}_A\ket{1}_B, \\
  \ket{\psi_{10}} &=& \ket{1}_A\ket{0}_B,
\end{eqnarray}
where the $\pm$ refers to bosons/fermions, respectively. Now, we label local measurement projectors as $\Pi_{a/b}^{A/B}=\frac{1}{2}\left(\openone+(-1)^{a/b}\sigma_{A/B}\right)$ for $A$ and $B$, respectively. Here,  $\sigma_{A/B}^2=\openone$  are binary observables. These operators reside in the qubit subspaces spanned by $\beta_A$ and $\beta_B$. { To implement this  measurements one can locally interfere the ``$M$'' and ``$S$'' particles at a beam splitter for each of the two locations. Notice however, that the particles on Bob's and Alice's sides will never be brought together again.}

Let us analyze the probability of success case by case. For $x=y=0$, in half of the cases $A$ and $B$ achieve $1/2$ (this accounts for the situations where local detectors register two or no particles). In the other two cases, the probability of success is given by $\bra{\psi_{00}}\Pi_{0}^{A}\otimes\Pi_0^B+\Pi_{1}^{A}\otimes\Pi_1^B\ket{\psi_{00}}$. Hence,one has to average these two possibilities:
\begin{eqnarray}\label{prob00}\nonumber
 p(00|00)+p(11|00)&=&\frac{1}{4}+\frac{1}{2}\bra{\psi_{00}}\Pi_{0}^{A}\otimes\Pi_0^B+
 \Pi_{1}^{A}\otimes\Pi_1^B\ket{\psi_{00}}\\
 &=&\frac{1}{2}+\frac{1}{4}\bra{\psi_{00}}\sigma_A\otimes\sigma_B\ket{\psi_{00}}.
\end{eqnarray}
Similarly, for $x=0$ and $y=1$, the measurement reveals one particle per party in $1/4$ of the cases only (in other $3/4$ of the cases $A$ and $B$ achieve the success of $1/2$ by outputting random results), thus we have
\begin{eqnarray}\label{prob01}\nonumber
 p(01|01)+p(10|01)&=&\frac{3}{8}+\frac{1}{4}\bra{\psi_{01}}\Pi_{0}^{A}\otimes\Pi_1^B+
 \Pi_{1}^{A}\otimes\Pi_0^B\ket{\psi_{01}}\\
 &=&\frac{1}{2}-\frac{1}{8}\bra{\psi_{01}}\sigma_A\otimes\sigma_B\ket{\psi_{01}}.
\end{eqnarray} 
In complete analogy, for $x=1$ and $y=0$ we get
\begin{eqnarray}\label{prob10}\nonumber
 p(01|10)+p(10|10)&=&\frac{3}{8}+\frac{1}{4}\bra{\psi_{10}}\Pi_{0}^{A}\otimes\Pi_1^B+
 \Pi_{1}^{A}\otimes\Pi_0^B\ket{\psi_{10}}\\
 &=&\frac{1}{2}-\frac{1}{8}\bra{\psi_{10}}\sigma_A\otimes\sigma_B\ket{\psi_{10}}. 
\end{eqnarray} 
Finally, for the case $x=y=1$, $A$ and $B$ output random bits always, thus, the probability of success is  $p(00|11)+p(11|11)=1/2$. Putting everything together we have
\begin{equation}\label{Pwin}
P_{win}=\frac{1}{2}\pm \frac{1}{32}\left(\bra{0}\sigma_A\ket{1}\bra{1}\sigma_B\ket{0}+
\bra{1}\sigma_A\ket{0}\bra{0}\sigma_B\ket{1}\right).
\end{equation}
{The maximal value is achieved for $\sigma_A= \pm \sigma_B=\sigma_x$, where $\pm$ refers again to bosons/fermions, respectively. Therefore, the optimal quantum value is $P_{win}^{Q1}=\frac{9}{16}$. We can also explicitly calculate all the coherence equalities, as defined in Eq.~\eqref{Iquant}, that for this choice of measurements give a maximal violation of $I_{ab}=\frac{(-1)^{a\oplus b}}{8}$.}

\textbf{\textit{Second example.}} {We introduce now an alternative example that, while not making use of pre-shared coherence, requires to violate the particle-number conservation, thereby making this example fundamentally unattainable for fermions \cite{aharonov2000nonlocal}.} Suppose that the initial state of the particles produced by the source is (in second quantization) $\ket{\psi}=s_0\ket{1}_A\ket{0}_B+s_1 \ket{0}_A\ket{1}_B$, with $|s_0|^2+|s_1|^2=1$. Here $\ket{0}_A$ and $\ket{1}_A$  are the states associated with, respectively, zero or one particle in mode $A$ (i.e., in the path $S-A$). The analogous notation holds for mode $B$. The associated two-mode density matrix is given by $\rho^{AB}=|\psi\rangle\langle\psi|$.
The operation associated to blocking the path is the ``blocking'' operator $\mathcal{B}$ \cite{Rozema19}. When a system in mode $A$, prepared in some state $\rho^A$, arrives at the blocker, it undergoes the operation $\mathcal{B}_A(\rho^A)=|0\rangle\langle0|_A$, for an arbitrary input state $\rho^A$. Adapted to the present scenario, we have the following blocking operators for modes $A$ and $B$, respectively:
\begin{equation}
\begin{array}{ll}
\mathcal{B}_A\left(\rho^{AB}\right) =  |0\rangle\langle0|_A\otimes  \rho^{B}, \vspace{0.2cm}\\
\mathcal{B}_{B}\left(\rho^{AB}\right) = \rho^{A}\otimes|0\rangle\langle0|_B,
 \end{array}
\end{equation}
where $\rho^{A}=Tr_B\left(\rho^{AB}\right)$ and $\rho^{B}=Tr_A\left(\rho^{AB}\right)$ are the reduced density matrices of the two subsystems $B$ and $A$ respectively. If no blocker is introduced, then the corresponding state does not undergo any influence, thus the identity transformation is applied. 

The introduction of the blockers or otherwise transforms the input state $\rho^{AB}$ into the state $\rho_{xy}$, which now encodes the inputs $x$ and $y$ (as defined above) by the transformation:
\begin{equation}
\rho_{xy}= (\mathcal{B}_A)^{x}(\mathcal{B}_B)^{y} \rho^{AB}.
\end{equation}
Hence, one has $\rho_{00}=\rho^{AB}$, $\rho_{01}=\rho_A\otimes|0\rangle\langle0|_B$, $\rho_{10}=|0\rangle\langle0|_A\otimes\rho_B$ and $\rho_{11}=|0\rangle\langle0|_A\otimes|0\rangle\langle0|_B$.
Let $A$ and $B$ perform binary measurements in their respective Fock spaces, defined by $\Pi_{a/b}^{A/B}=\frac{1}{2}\left(\openone+(-1)^{a/b}\sigma_{A/B}\right)$. Here $\sigma_{A/B}$ are single-quibit operators that reside in local Fock spaces spanned by the vacuum $\ket{0}_{A/B}$ and single-particle state $\ket{1}_{A/B}$. The conditional probabilities are given by $p(ab|xy) = \mathrm{Tr}\left[\rho_{xy} \Pi_a^A \otimes \Pi_b^B\right]$ and 
\begin{eqnarray}
I_{ab}&=&\mathrm{Tr}\left[\Pi_a^A \otimes \Pi_b^B\sum_{xy=0}^1(-1)^{x\oplus y}\rho_{xy}\right]\\
&=&s_0 s_1^{*}\bra{0}\Pi_a^A\ket{1}\bra{1}\Pi_b^B\ket{0}+h.c.
\end{eqnarray}
The probability of success \eqref{classicalwinprob} evaluates to
\begin{equation}
P_{win}=\frac{1}{2}+\frac{1}{8}(s_0 s_1^{*}\bra{0}\sigma_A\ket{1}\bra{1}\sigma_B\ket{0}+h.c.).
\end{equation}
The maximum is achieved for $\sigma_A=\sigma_B=\sigma_x$ and $s_0=s_1=\frac{1}{\sqrt2}$, for which we find: 
\begin{equation}
P^{Q_2}_{win}=\frac{5}{8}.
\end{equation}
One  can then evaluate  $I_{ab}=\frac{(-1)^{a\oplus b}}{4}$, which clearly violates all four coherence equalities \eqref{Belleq}.

Note that in this example the implementations of measurements $\Pi_{a/b}^{A/B}$ requires the read-out in superposition of the vacuum state and a single-particle excitation, thus requiring the violation of the particle number conservation. This makes the proposal demanding for bosonic particles, whereas the parity superselection rule completely forbids this for fermions \cite{Wightman1995, aharonov2000nonlocal}. Nevertheless, such measurements are in principle physical for bosonic particles, and  have, in fact, been implemented for single-photons by transferring the photonic excitations to atoms \cite{van2005single,fuwa2015experimental,Rosenfeld17}.

\section{Conclusions}
In this paper, we have investigated the possibility of witnessing a quantum superposition of communication, by means of probabilistic correlations between distant parties. Remarkably, this procedure does not require to recombine the beams to unambiguously detect a superposition state. Phrasing this problem in terms of a communication game, we have derived a coherence equality that is satisfied by any classical communication (i.e., when the information carrier is a well-localized particle). On the contrary, we have provided two concrete examples --experimentally implementable-- where the use of quantum resources for communication allows a violation of the coherence equality, and therefore certifies the presence of quantum superposition.
Remarkably, this is done by local measurements only.

\begin{acknowledgements}
\textbf{Acknowledgments - }We would like to thank Dengke Qu and Valerio Scarani for useful comments that greatly improved our manuscript. B.D. acknowledges support from an ESQ Discovery Grant of the Austrian Academy of Sciences ({\"O}AW) and the Austrian Science Fund (FWF) through BeyondC (F71). F.D.S. acknowledges the financial support through a DOC Fellowship of the
Austrian Academy of Sciences ({\"O}AW).
\end{acknowledgements}

\bibliographystyle{achemso}
\bibliography{library1}

\clearpage
\section*{Appendix A}
\subsection{Statistical test of the coherent equality}
One could wonder how is the classical equality $P_{win}^{Class}=\frac{1}{2}$ testable in practice, as being only satisfied for one precise value instead of an interval (standardly one tests the Bell's inequality, i.e. $P_{win}\leq P_{Class}$). In order to make an operationally meaningful statement, suppose an experimenter has performed $N$ repeated game trials, as defined above, and has recorded the sequence of outcomes $f_i$ ( for $i=1 \cdots N$), where $f_i=1$ means success and $f_i=0$ failure of the $i$-th trial. The estimation of the probability of success is given by the relative frequency $F_N=\frac{1}{N}(f_1+\cdots + f_N)$. It is clear that the classical equality $P_{win}^{Class}=1/2$ guarantees $F_N$ to be a martingale random variable, therefore by the  straightforward application of the Azuma's theorem \cite{azuma1967weighted} we have
\begin{equation}
 P\left[\left|F_N-\frac{1}{2}\right|\geq \varepsilon\right]\leq 2 e^{-2N\varepsilon^2},
\end{equation}
for every $\varepsilon>0$. This means that the probability of deviation of the relative frequency from $1/2$ is exponentially suppressed with the number of experimental trials for any possible classical scenario. Finally, from a concrete data record one can use the inequality above to calculate the statistical significance of a (possible) violation. 
\end{document}